\documentclass[twocolumn,aps,preprintnumbers,superscriptaddress]{revtex4}
\usepackage{graphicx}
\usepackage{bm}
\usepackage{amsmath, amsthm, amssymb}

\begin{document}
\title{Broadband Control on Scattering Events with Interferometric Coherent Waves}
\author{Jeng Yi Lee}
\affiliation{
Department of Opto-Electronic Engineering, National Dong Hwa University, Hualien 974, Taiwan }

 \author{Lujun Huang}
\affiliation{School of Engineering and Information Technology, University of New South Wales, Canberra ACT 2600, Australia}

\author{Lei Xu}
\affiliation{School of Engineering and Information Technology, University of New South Wales, Canberra ACT 2600, Australia}

\author{Andrey E. Miroshnichenko}
\affiliation{School of Engineering and Information Technology, University of New South Wales, Canberra ACT 2600, Australia}

\author{Ray-Kuang Lee}
\affiliation{
Institute of Photonics Technologies, National Tsing Hua University, Hsinchu 300, Taiwan}
\affiliation{Physics Division, National Center of Theoretical Science, Hsinchu 300, Taiwan}
\date{\today}

\begin{abstract} 
We propose a universal strategy to realize a broadband control on arbitrary scatterers, through multiple coherent beams.
By engineering the phases and amplitudes of incident beams, one can suppress the dominant scattering partial waves, making the obstacle lose its intrinsic responses in a broadband spectrum.
The associated coherent beams generate a finite and static region, inside which the corresponding electric field intensity and Poynting vector vanish.
As a solution to go beyond the sum-rule limit, our methodology is also irrespective of inherent system properties, as well as extrinsic operating wavelength, providing a non-invasive control on the wave-obstacles interaction for any kinds of shape.
\end{abstract}

\maketitle
Making functional sub-wavelength scatterers has been attractive  for a variety of applications, such as superdirective scatterers~\cite{super1,super2,super3}, perfect absorption objects~\cite{absorber1,absorber2},  magnetic resonantor based devices~\cite{magnetic1,magnetic2,magnetic3}, Kerker effect and beyond \cite{kerker1,kerker2,kerker3}, anapole~\cite{anapole}, and superscattered objects ~\cite{superscattering1, superscattering2}. In particular,  to have invisible cloaks, the concepts of transformation optics \cite{controlling,conformal} and scattering cancellation method~\cite{kerker, alu1,alu2} have been applied not only to electromagnetic waves, but also to acoustic~\cite{acoustic1,acoustic2,acoustic3,sc1,sc2} and water waves~\cite{fluid1,fluid2}, thermal diffusion science \cite{thermal1,thermal2,sc8}, quantum matter waves~\cite{quantum1,quantum2, quantum3,sc6,sc7}, and elastic wave in solids~\cite{elastic1,elastic2,elastic3}.
However, for these methods and the consequently improved efforts \cite{carpet}, we still suffer from the superluminal propagation \cite{quest1,quest2}, and limited operating bandwidth imposed by Kramers-Kronig relation~\cite{quest3,quest4}.
To manipulate light-obstacles interaction in nanoscales, it is still desirable to have a non-invasive and efficient way to have objects working in a broadband spectrum.

As pointed out by  E. M. Purcell \cite{sum}, the integration of the extinction cross section over all the spectra is related to the static electric and magnetic material parameters, leading to the sum-rule limit.
Therefore, under a plane wave excitation, no scattering systems can remain stationary scattering responses in a broadband spectrum.
In this Letter, we demonstrate that it is possible to turn off or amplify the target scattering partial waves with interferometric coherent waves.
With a proper setting on the phases and intensities~\cite{linear1}, destructive interferometry on the dominant partial waves can be achieved, resulting in an arbitrary object invisible.
At the same time,  a finite and static region emerges, inside which the electric field and the corresponding Poynting vector almost completely vanish.
Instead, high-order scattering partial waves are excited even the physical size of the scatterer  is smaller than the size of this finite region.
Moreover, the operating wavelength for such a destructive removal of excitation exists for a broadband spectrum, overcoming the fundamental sum-rule limit obtained from a single plane wave excitation.
Also, there exist more than one settings for the interferometric field in the excitation, demonstrating the flexibility for experimental implementations.
The robustness of our methodology on invisibility is also verified by introducing the deviations in the scatter displacements, intensities and illumination angles of incident beams, and different shapes of systems. 
Our results pave an alternative route to manipulate waves and obstacles in the extremely small scale.

\begin{figure*} 
\includegraphics[width=18.0cm]{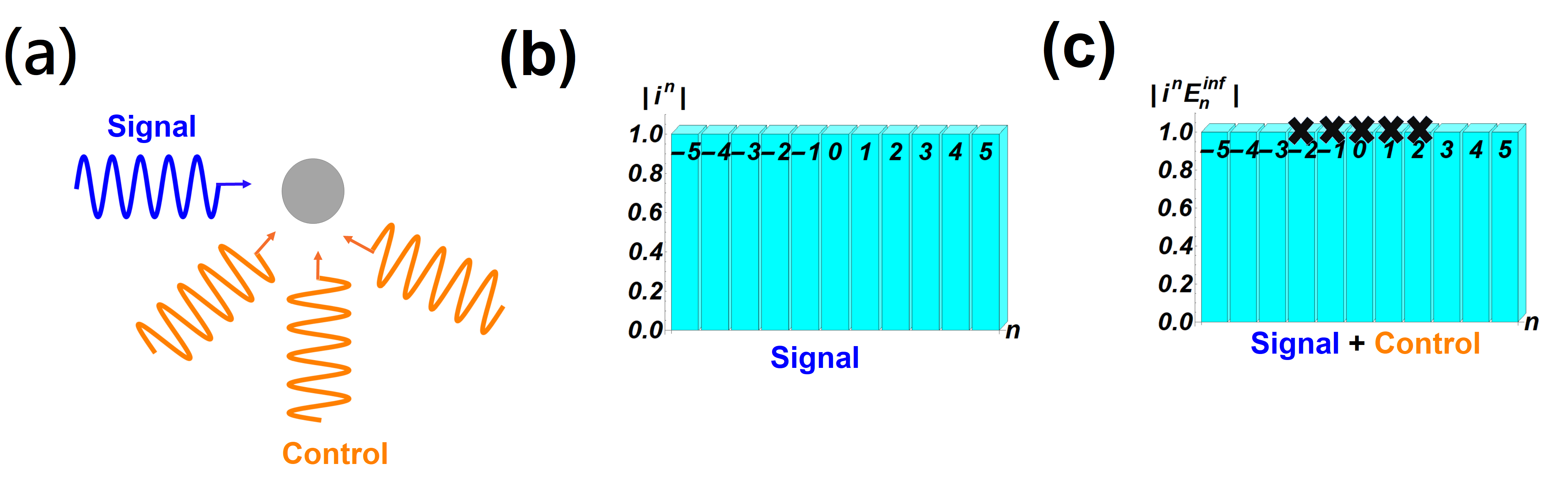}
\caption{(a) A schematic of a cylindrical scatterer, which is irradiated by the interferometric signal and control waves. 
For a single plane wave (signal) excitation, the scattering strength on each partial wave $\vert i^{n}\vert$ is shown in (b); while for multiple waves (single $+$ control) excitation, one can turn off the initially dominant scattering events, marked by the {\bf $\times$}-sign in (c).}
\end{figure*}

Without loss of generality, we consider the illumination waves composed by a set of plane waves on a cylindrical scatterer, as illustrated in Fig. 1(a).
Here, the symmetrical axis of the cylindrical axis is chosen as the $z$-axis.
For a single plane wave of \textbf{s}-polarized electric field propagating in $x$-axis, it can be described as   $E_{1}\hat{z}e^{ik_0r\cos\theta}$ with the signal wave denoted by $E_1$, the environmental wavenumber  $k_0$, and the azimuthal angle $\theta$.
By a combination of proper eigenstates $\nu_n(\vec{r})$, which rely on the scatterer structures, we have $E_{1}\hat{z}e^{ik_0r\cos\theta}=\hat{z}E_1\sum_{n}\phi_n\nu_n(\vec{r})$, with a complex coefficient $\phi_n$.
The time dependence for each plane wave has the form $e^{-i\omega t}$.
As an example for the cylindrical scatter, we adopt the Bessel and the first kind of Hankel functions obeying the Helmholtz equation for the eigenstates, i.e.,  $J_{n}(k_0r)e^{in\theta}$ and  $H^{(1)}_{n}(k_0r)e^{in\theta}$, respectively.
Then, the incident wave, denoted as {\it signal}, can be expressed as $E_{1}\hat{z}e^{ik_0r\cos\theta}=E_1\hat{z}\sum_{n=-\infty}^{\infty}i^nJ_{n}(k_0r)e^{in\theta}$, here the index, $n$, represents a series of partial waves~\cite{book1}.
The associated scattering wave generated by the scatterer has the form 
$\vec{E}_{sc}=E_{1}\hat{z}\sum_{n=-\infty}^{\infty}i^{n}a_{n}^{TE}H_{n}^{(1)}(k_0r)e^{in\theta}$, with the complex scattering coefficient $a_n^{TE}$.
The factor $i^{n}$ indicates the excitation strength in each partial wave.

For the excitation of a signal plane wave, the resulting excitation strength from this cylinder can be expressed by $\vert i^{n}\vert$, as shown in Fig. 1(b). 
However, as we are going to illustrate, our target is to demonstrate that the irradiation from a proper setting of signal and control waves, can turn off the initially dominant scattering events,  as illustrated in Fig. 1(c), resulting in scatterers lose their functionality.

To describe the total illumination and  scattering waves, one has
$\vec{E}_{in}=\hat{z}\sum_{n=-\infty}^{\infty}i^{n}J_{n}(k_0r)e^{in\theta}E_n^{inf}$ and
$\vec{E}_{sc}=\hat{z}\sum_{n=-\infty}^{\infty}i^nE_n^{inf}e^{in\theta}H_n^{(1)}(k_0r)a_n^{TE}$, with the introduction of an interfering factor $E_n^{inf}$, which has the form
\begin{eqnarray}
E_n^{inf}=E_1+\sum_{m=2}^{m=s}e^{-in\Phi_m}E_m.
\end{eqnarray}
Here, the first term in the right-handed side of Eq. (1) corresponds to the signal wave; while the others represent $s-1$ ($s\geq 2$) control waves whose complex wave amplitudes and incident angle are defined as $E_m$ and $\Phi_m$, respectively.
The corresponding scattering and absorption powers are
$P_{sc}=2/k_0\times\sqrt{\epsilon_0/\mu_0}\sum_{n=-\infty}^{\infty}\vert E_n^{inf}\vert^2\vert a_n^{TE}\vert^2$ and 
$P_{abs}=-2/k_0\times\sqrt{\epsilon_0/\mu_0}\sum_{n=-\infty}^{\infty}\vert E_n^{inf}\vert^2[Re(a_n^{TE})+\vert a_n^{TE}\vert^2]$,
with the environmental permittivity and permeability denoted as  $\epsilon_0$ and $\mu_0$, respectively.

Now, suppose that our scattering system has $2N+1$  dominant partial waves (scattering channels).
The only way to eliminate the scattering of these dominant partial waves is to produce the destructive interference of these target channels, i.e., $E_n^{inf}=0$ from $n=[-N, N]$.
However, it can be proved straightforwardly that a total excitation of $2N+1$ irradiation waves (including the signal wave), only leads to a trivial zero solution \cite{linear1}.
To obtain a non-trivial solution, one possibility is to expand the amount of control waves to $2N+1$ in total at least. Then,  we have the following $2N+1$ equations to be satisfied:
\begin{eqnarray}
E_1+\sum_{m=2}^{2N+2}e^{-in\Phi_m}E_m = 0,\quad \text{for} \quad n=[-2N,2N].
\end{eqnarray}
Here, in each equation there are three degrees of freedom for the extrinsic control parameters: intensity and phase of a control wave $E_i$, and the corresponding incident angle $\Phi_i$.
In general, one should have a variety of solutions to satisfy the necessary condition in Eq. (2).

 \begin{figure*}
\includegraphics[width=16.0cm]{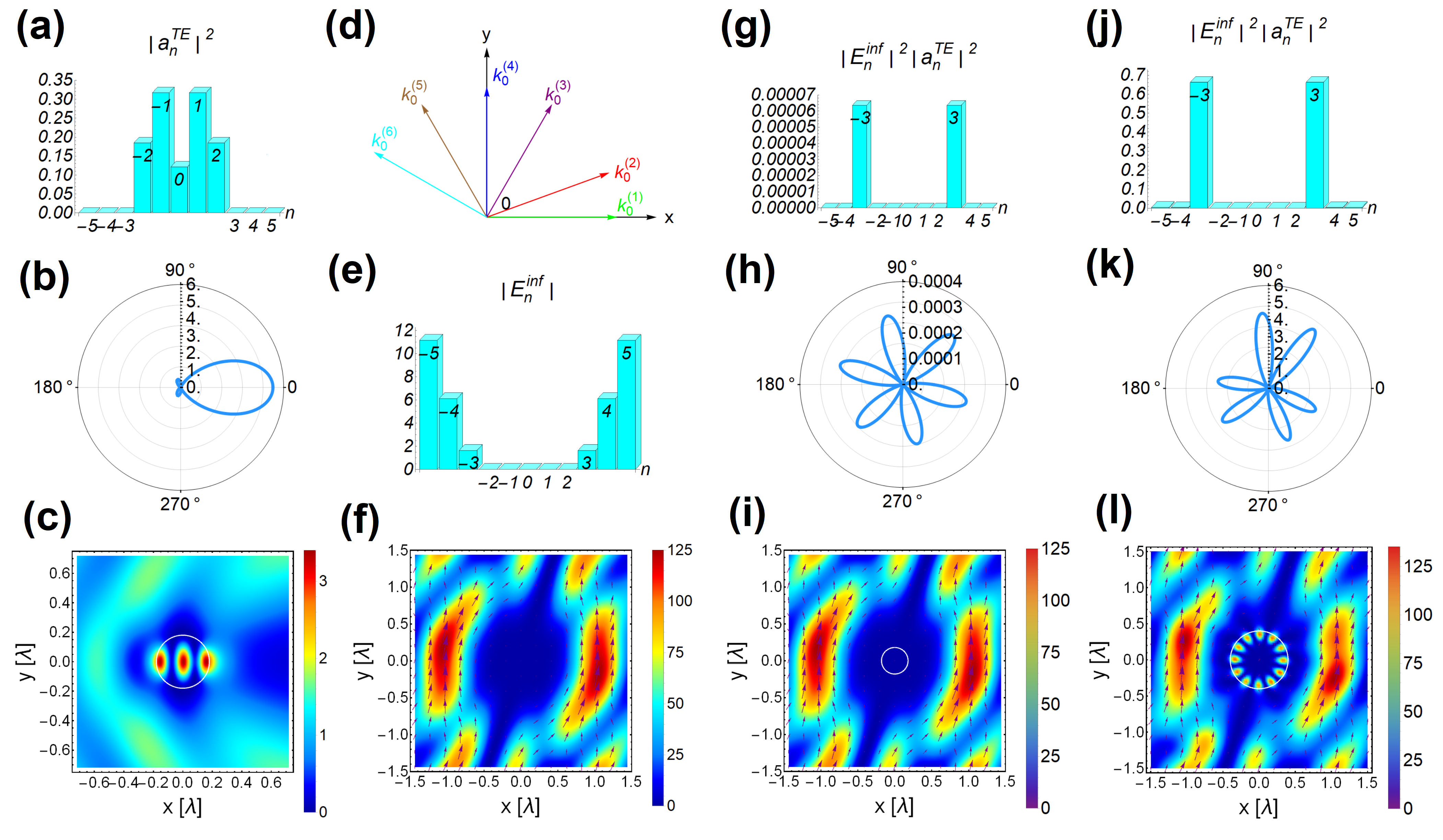}
\caption{For a single wave (signal) excitation: (a) The resulting scattering strength $\vert a_n^{TE}\vert$ is revealed for each partial wave. (b) The corresponding far-field distribution and (c) the intensity of the electric field.
For a multiple (signal $+$ control) excitations: (d) The illumination configuration is depicted for five control waves.
Here, $k_0^{(1)}$ denotes the signal wave; while  $k_0^{(i)}$, $i=2, \dots 6$ denote the control waves. 
The corresponding magnitudes of the interfering factor for each channel are depicted in (e).
The electric field intensity and Poynting vector from the multiple coherent waves are shown in (f), as the background.
With the illumination configuration in (d),  for a smaller radius $a = 0.18\lambda$, the associated partial scattering powers by the cylindrical scatterer (marked by the circle in White-color) is depicted in (g); while the corresponding far-field pattern and electric field are revealed in (h) and (i), respectively.
Instead, for a larger radius $a = 0.4\lambda$, the associated partial scattering powers, corresponding far-field pattern and electric field are revealed in (j), (k), and (l), respectively.
Note that the values given in (j) are three orders of magnitude large. Here, only $n=\pm 3$rd scattering channels are excited and amplified.}
\end{figure*}

To demonstrate our control on the scattering events,  a silicon-embedded system is considered, such as silicon embedded with a high refractive index $\epsilon_1=12$~\cite{handbook}.
Here, we tackle the first five dominant scattering channels, as shown in Fig. 2(a).
These five scattering events correspond to the electric dipole ($n=0$), magnetic dipole ($n \pm 1$), and magnetic quadrupole ($n \pm 2$).
The corresponding far-field scattering distribution and the intensity of the electric field are illustrated in Figs. 2(b) and 2(c) for a single wave excitation (signal only).
Details on how to obtain the far-field scattering distribution are provided in Supplementary Materials. 
Now, in order to suppress these five dominant scattering partial waves, we construct an illumination system with another five control waves, denoted as $(E_2, E_3, E_4, E_5, E_6)$, with  the corresponding incident angles $(\Phi_2,\Phi_3,\Phi_4,\Phi_5,\Phi_6)$. 
Then, we rewrite Eq. (1) into  the following matrix presentation:
\begin{equation}
\begin{split}
\begin{bmatrix}
-1\\
-1\\
-1\\
-1\\
-1\\
\end{bmatrix}=
\begin{bmatrix}
e^{-2i\Phi_2} & e^{-2i\Phi_3} &e^{-2i\Phi_4} &e^{-2i\Phi_5} & e^{-2i\Phi_6} \\
e^{-i\Phi_2} & e^{-i\Phi_3} &e^{-i\Phi_4} &e^{-i\Phi_5} & e^{-i\Phi_6} \\
1 & 1 & 1 & 1 &1\\
e^{i\Phi_2} & e^{i\Phi_3} &e^{i\Phi_4} &e^{i\Phi_5} & e^{i\Phi_6} \\
e^{2i\Phi_2} & e^{2i\Phi_3} &e^{2i\Phi_4} &e^{2i\Phi_5} & e^{2i\Phi_6} \\
\end{bmatrix}
\begin{bmatrix}
E_2\\
E_3\\
E_4\\
E_5\\
E_6\\
\end{bmatrix}.
\end{split}
\end{equation}
Here, without loss of generality, we set $E_1=1$.
As  indicated in Fig. 2(d), we also chose the  illumination angles as $[\Phi_2=\pi/9$, $\Phi_3=\pi/3$, $\Phi_4=\pi/2$, $\Phi_5=2\pi/3$, $\Phi_6=5\pi/6]$, based on which one can obtain the corresponding control wave amplitudes by solving Eq. (3), resulting in  $[E_2=-2.17$, $E_3=3.28$, $E_4=-3.78$, $E_5=2.31$, $E_6=-0.64]$.
In principle, one can set the incident angles arbitrarily and find out the corresponding complex amplitudes by solving Eq. (3).
With these obtained results, we analyze the interfering factors for each excited scattering events, as shown in   Fig. 2(e).
As one can see, a complete destructive interference condition happens for the target channels $n=[-2,-1,0,1,2]$, with all the zero values. 
Meanwhile, non-zero interfering factors emerge on non-target scattering channels, i.e., $n= \pm 3, \pm 4$, and $\pm 5$.
This result indicates that when the destructive interferometry applies to the dominant scattering channels, one can completely suppress the scattering events at the price that the originally non-dominant scattering channels are amplified.
With the comparison between single and multiple excitations, Fig. 2(c) and 2(f), the intensity of electric fields, as well as the energy Poynting vectors, are totally different.

With the same set of illumination configuration given in Fig. 2(d), we reveal another extreme scenario only with $n = \pm 3$rd channels supported, as shown in Fig. 2(g). 
Now, one can easily see that the corresponding far-field scattering pattern shown in Fig. 2(h) is significantly suppressed, i.e., at least three orders of magnitude smaller.
The resulting electric field, as well as the time-averaged Poynting vectors, shown in Fig. 2(i) clearly demonstrate that the energy bypasses scatterer in the central region.
Moreover, a finite and static region emerge within $x=[-0.5\lambda,0.5\lambda]$ and $y=[-0.5\lambda,0.5\lambda]$, inside which nearly all the intensity and energy Poynting vectors vanish.
Here, $\lambda$ is the wavelength of illumination waves.
With the comparison between Figs. 2(f) and 2(i), it is almost indistinguishable both for the field distribution and Poynting vectors, supporting the realization of invisibility.

At a quick glance, as the existence of a finite and static region induced by the multiple wave excitation, one may contribute it for the reason to make the scatterer lose its functionality, as the physical size of our scatterers is smaller than the size of this zero-field region.
As shown in Fig. 2(e), even though the interfering factors are completely suppressed for $n=-2, -1, 0, 1, 2$, other partial waves still survive from the wave-obstacle interaction.
To highlight the size effect,  we choose a bigger scatterer by changing the radius of our cylinder from $a=0.18\lambda$ to $a = 0.4\lambda$.
With the same setting in Fig. 2(d), now, the resulting scattering events are enhanced for the $n= \pm 3$rd scattering channels, as shown in Fig. 2 (j).
Nevertheless, with different far-field scattering pattern and field intensity (also the Poynting vector), as shown in Figs. 2(k) and 2(l), respectively, the invisibility is broken.

 \begin{figure}
\includegraphics[width=8.5cm]{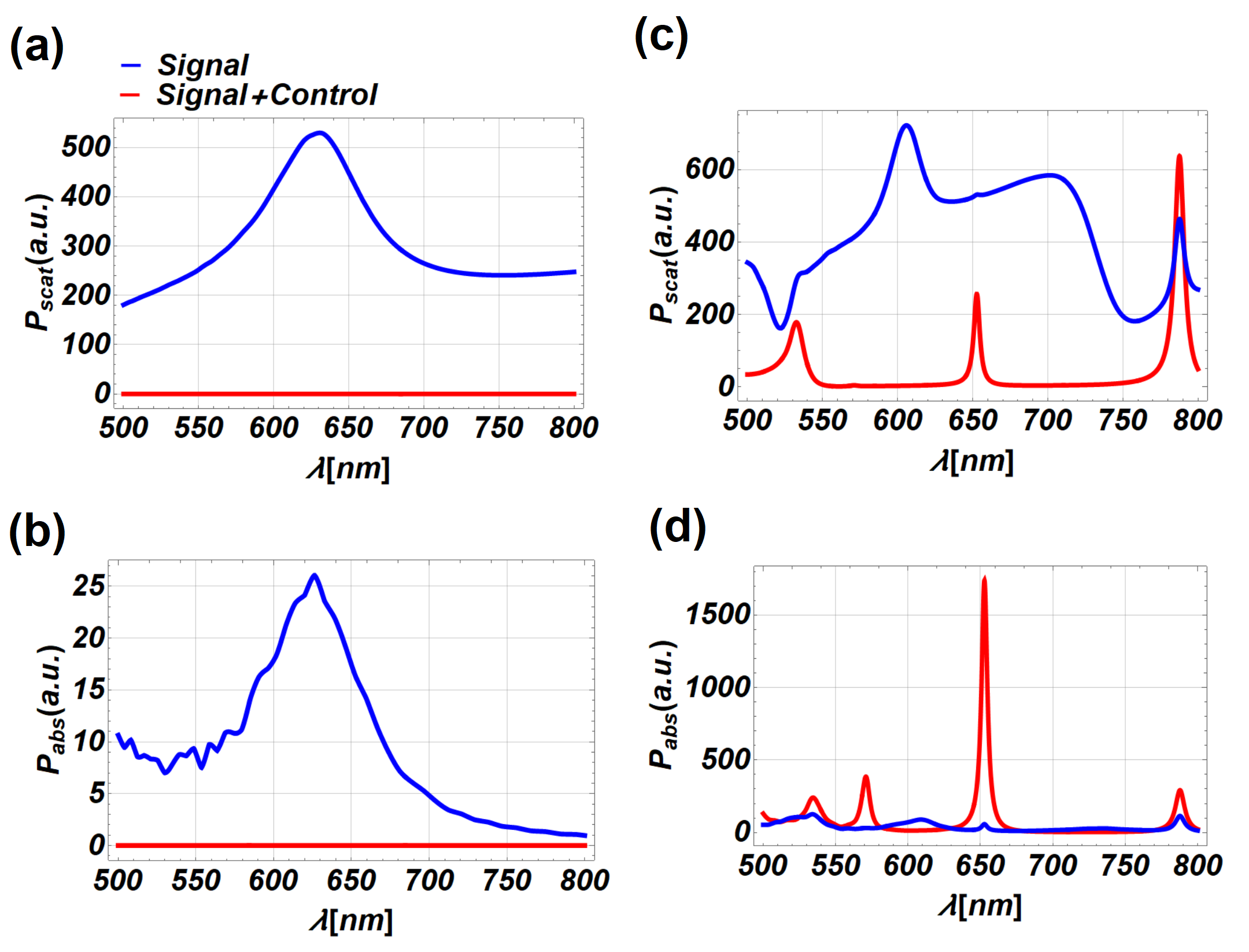}
\caption{ The scattering (a, c) and absorption power (b, d) spectra under the excitation of a single plane wave (signal) and multiple coherent waves (single $+$ control), depicted in Blue- and Red-colors, respectively. Here, a smaller radius of cylinder, $a=60$nm is considered in (a-b); while a larger radius, $a=170$nm is considered in (c-d).}
\end{figure}

Instead of using a structured wave to select the multipolar modes~\cite{linear1,linear2,linear3,linear4}, our approach with multiple wave excitation is entirely different.
Furthermore, our methodology can support broadband control through the interferometric coherent waves.
If we keep all the system parameters fixed, including the illumination angles, intensities and phases of control waves, but only tune the incident wavelength.
For a smaller size of the scatterer, in Fig. 3 (a-b), we set the silicon cylinder with the radius $a=60$nm and scan the incident  wavelength from $500$ to $800$nm.
Interestingly, compared to the plane wave excitation, depicted in Blue-color, by multiple wave excitation, both the scattering and absorption power spectra give us the zero values in this wavelength range, as depicted in Red-color.
Even though it is known that for any invisible cloak illuminated by a single plane wave, Kramers-Kronig relation and sum-rule limit prevent the realization of a broadband operation.
Our results demonstrate the scenario to go beyond the sum-rule limit,  indicating the scatterer system working at this wavelength window with lowest-orders in the partial scattering waves.

\begin{figure}[t]
\includegraphics[width=8.4cm]{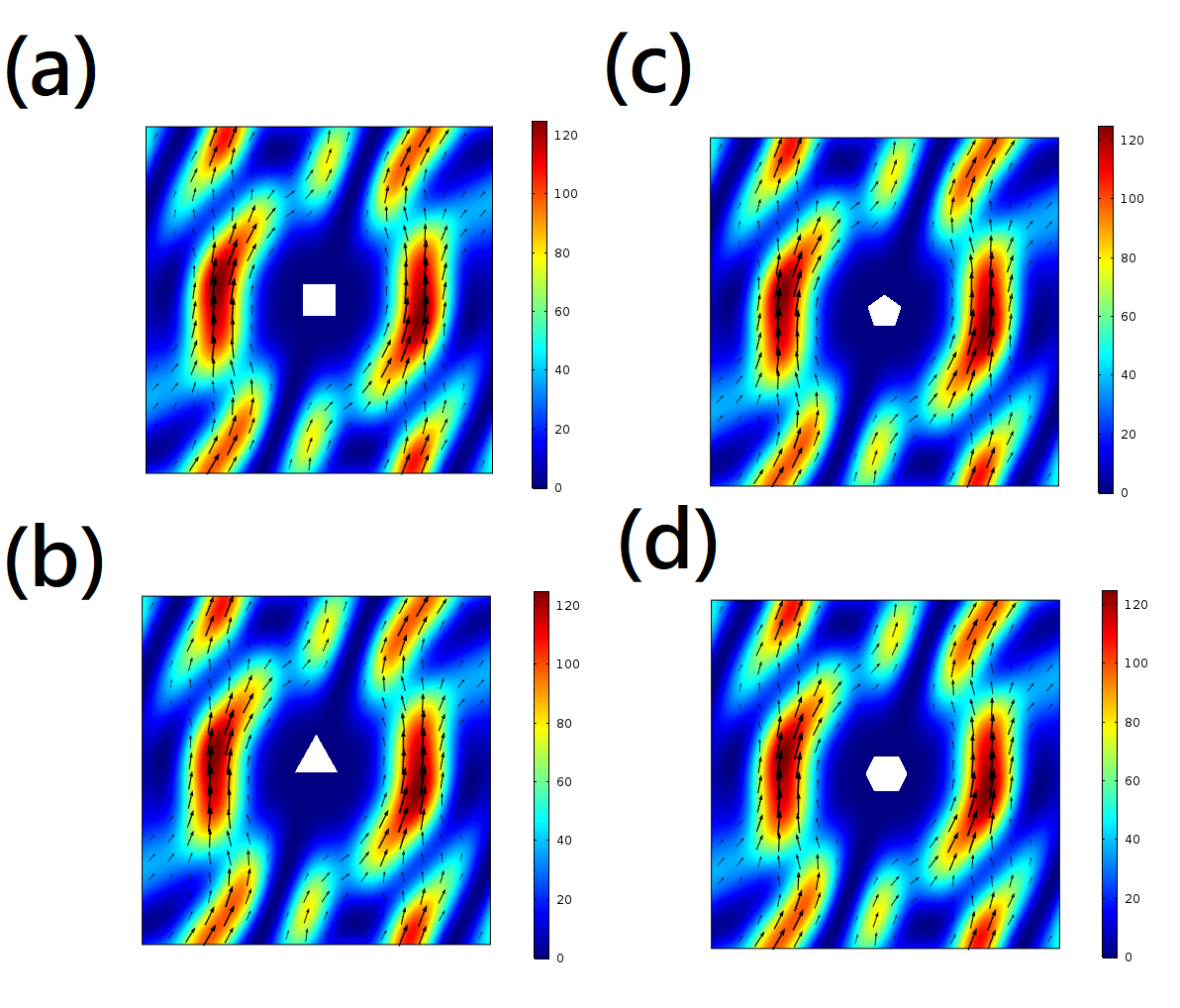}
\caption{With the same illumination configuration in Fig. 2(d), when one places the scatterers in different shapes: (a) square, (b) triangle, (c) pentagon, and (d) hexagon, the resulting electric field distribution and Poynting vectors all remain unchanged.}
\end{figure}

For a larger size silicon cylinder, in Fig. 3(c-d), we set $a=170$nm as an example.
It is known that as the size of the scatterer increases, the scattering power by the signal plane wave increases; while the corresponding absorption power decreases. 
Even though such a larger size system can support higher-order scattering channels, the target scattering events can remain suppressed with a multiple wave excitation.
As a guideline, one solution to further suppress these higher-order scattering channels is to introduce more control waves.

As for the influences on the invisibility from the mismatches in intensities and illumination angles of incident beams, as well as the scatterer displacement, one can apply Graft's addition theorem~\cite{addition} to have a systematic study. In Supplementary Materials, a detailed analysis is presented.
Our finding reveals that the interferometric method is robust to allow these mismatching, offering flexibility toward the experimental implementation. 
For different kinds of shape,  we also studied scatterers in the shape of a square, triangle, hexagon, and a pentagon, by Comsol\cite{comsol}, as shown in Fig. 4.
All of our outcomes can support invisibility.
Even though the analysis in this work is demonstrated for the two-dimensional system, but it is readily applied to a three-dimensional scatterer or clusters.
Last but not least, we note that the invisibility or enhanced scattering of the wave does not depend on individual parameters, e.g., size, structures, or materials.

In summary, we have demonstrated a novel way by extrinsically imposing interferometric multiple waves to manage the excitation of partial waves.
Compared to the single plane wave illumination, we reveal the possibility to support invisibility and to enhance target scattering partial waves, irrespective of internal system configuration.
Unlike the known wave-obstacle interaction, which strongly relies on the material dispersion, such a multiple wave illumination provides a non-invasion way to avoid this physical constraint.
It is the interferometric coherent waves, to support the existence of stationary scattering response for a  broadband wavelength, beyond the sum-rule limit.
The coherent control paves a new and exciting way to manipulate wave-obstacle interaction in the deep subwavelength scale for a variety of waves physics.

\noindent \textit{Acknowledgments:} This work is supported by Ministry of Science and Technology, Taiwan (107-2112-M-259-007-MY3 and 105-2628-M-007-003-MY4). The work of AEM was supported by the Australian Research Council and UNSW Scientia Fellowship.

\section{Appendix A: Formula for far-field scattering distribution}
Here, we give the formula for far-field scattering distribution. Start with 
\begin{equation}
P_{sc}(\theta)=\frac{1}{2}Re[\vec{E}_{sc}\times\vec{H}_{sc}^{*}]\cdot\hat{r}r,
\end{equation}
and apply the asymptotic analysis for the first kind of Hankel function, then, we can have
\begin{equation}
\begin{split}
\vec{E}_{sc}\sim\sqrt{\frac{2}{\pi k_{0}r}}e^{i(k_{0}r-\frac{\pi}{4})}\hat{z}\sum_{n=-\infty}^{\infty}i^{n}e^{in\theta}E_n^{inf}a_n^{TE}e^{-i\frac{n\pi}{2}},\\
(\vec{H}_{sc})_{\theta}\hat{\theta}\sim-\frac{k_{0}}{\omega\mu_0}\sqrt{\frac{2}{\pi k_{0}r}}e^{i(k_{0}r-\frac{\pi}{4})}\hat{\theta}\sum_{n=-\infty}^{\infty}i^{n}e^{in\theta}E_n^{inf}a_n^{TE}e^{-i\frac{n\pi}{2}}.
\end{split}
\end{equation}
Here, for magnetic fields, we only consider the $\hat{\theta}$ component, because only this term makes a contribution to radiation. Then, one can have
\begin{equation}
P_{sc}(\theta)=\frac{1}{\pi k_{0}}\sqrt{\frac{\epsilon_0}{\mu_0}}\vert \sum_{n=-\infty}^{\infty}e^{in\theta}E_n^{inf}a_n^{TE}\vert^2.
\end{equation}

\begin{figure}
\centering
\includegraphics[width=0.35\textwidth]{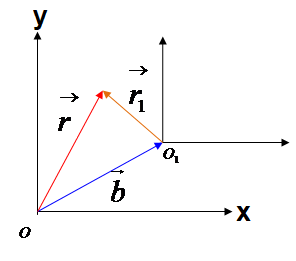}
\caption{Two coordinates denoted as $O$ and $O_{1}$, with a relative position vector, $\vec{b}$.}
\end{figure}

\section{Appendix B: Graft's addition theorem}
 
For fields expressed in different coordinates, we can apply Graft's addition theorem \cite{addition}. As shown in Fig. 5 for two coordinates, the  addition theorem indicates that waves expressed in the coordinate $O$ can be transformed into  the coordinate $O_1$. For our multiple wave irradiation at a  different coordinate $(r_1,\theta_1)$, we have
 \begin{equation}
 \begin{split}
  &\vec{E}_{in}(r_1,\theta_1)\\
  &=\hat{z}\sum_{m=-\infty}^{\infty} i^mJ_m(k_0r)e^{im\theta}E_m^{inf},\\
  &=\hat{z}\sum_{m=-\infty}^{\infty}i^m E_m^{inf}\sum_{n=-\infty}^{\infty} J_{m-n}(kb)e^{i(m-n)\beta}J_n(kr_1)e^{in\theta_1},\\
  &=\hat{z}\sum_{n=-\infty}^{\infty}J_n(kr_1)e^{in\theta_1}\sum_{m=-\infty}^{\infty}i^{m}E_m^{inf}J_{m-n}(kb)e^{i(m-n)\beta},\\
  &=\hat{z}\sum_{n=-\infty}^{\infty}J_n(kr_1)e^{in\theta_1}E_{n,g}^{inf},
 \end{split}
 \end{equation}
 where $(b,\beta)$ denotes the orientation of the new coordinate with respect to origin one, with $\vert\vec{b}\vert=b$ and $\beta$ be the  angle with respect to x-axis.
 Here, we can define a general interfering factor as
 \begin{equation}
 E_{n,g}^{inf}=\sum_{m=-\infty}^{\infty}i^mE_m^{inf}J_{m-n}(kb)e^{i(m-n)\beta},
 \end{equation}
 and the corresponding multi-beam irradiation as,
 \begin{equation}
 E_m^{inf}=\sum_{d=1}^{d=N}e^{-im\Phi_d}E_{d},
 \end{equation}
 supposing there are $N$ illumination waves.

Now, our interest is to find the corresponding scattering pattern when the object is placed at origin of the coordinate $O_1$. That is
\begin{equation}
\begin{split}
\vec{E}_{in}(r_1,\theta_1)&=\hat{z}\sum_nJ_n(kr_1)e^{in\theta_1}E_{n,g}^{inf},\\
\vec{E}_{sc}(r_1,\theta_1)&=\hat{z}\sum_nH^{(1)}_n(kr_1)e^{in\theta_1}E_{n,g}^{inf}a_n^{TE},\\
[\vec{H}_{sc}(r_1,\theta_1)]_{\theta_1}\hat{\theta_1}&\rightarrow\hat{\theta_1}\frac{1}{-i\omega\mu_0}\sum_n kH^{(1)'}_n(kr_1)e^{in\theta_1}E_{n,g}^{inf}a_n^{TE}.\\
\end{split}
\end{equation}

At the far-field region, by  applying the asymptotic analysis to the special functions, we have
\begin{equation}
\begin{split}
\vec{E}_{sc}(r_1,\theta_1)&\sim\hat{z}\sum_n\sqrt{\frac{2}{\pi k_0r_1}}e^{i(k_0r_1-\frac{\pi n}{2}-\frac{\pi}{4})}e^{in\theta_1}E_{n,g}^{inf}a_n^{TE},\\
\vec{H}_{sc}(r_1,\theta_1)&\sim-\frac{1}{\omega\mu_0}\hat{\theta_1}\sum_nk_0\sqrt{\frac{2}{\pi k_0r_1}}e^{i(k_0r_1-\frac{\pi n}{2}-\frac{\pi}{4})}e^{in\theta_1}E_{n,g}^{inf}a_n^{TE}.
\end{split}
\end{equation}

So the scattering power distribution when placed the antenna at this new location $O_1$ becomes
\begin{equation}
\begin{split}
P_{sc}(\theta_1)=\frac{1}{\pi k_0}\sqrt{\frac{\epsilon_0}{\mu_0}}\vert \sum_n e^{in\theta_1}e^{-i\frac{n\pi}{2}}E_{n,g}^{inf}a_n^{TE}\vert^2.
\end{split}
\end{equation}

  \begin{figure}
  \centering
\includegraphics[width=8.5cm]{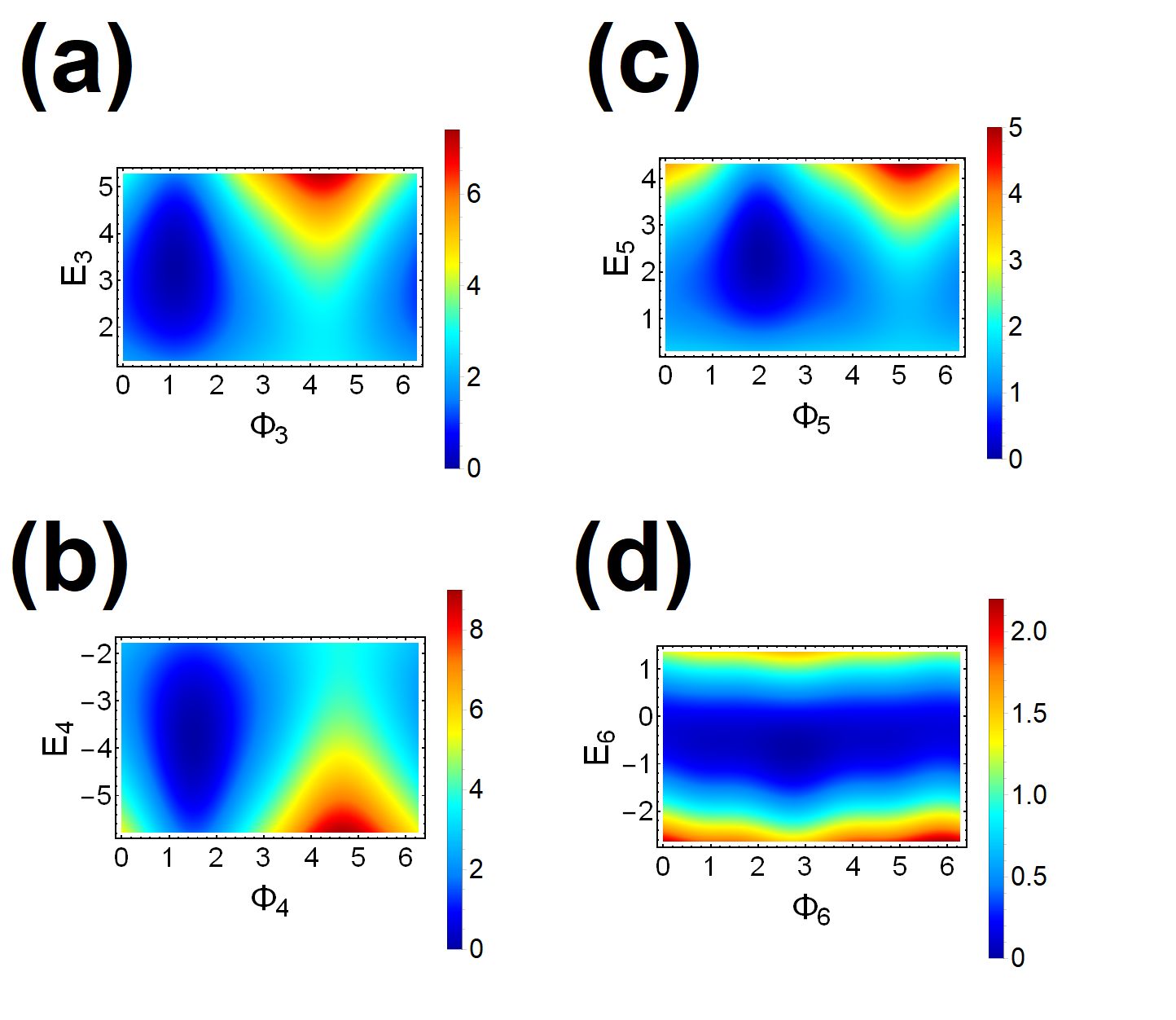}
\caption{Contour plots for the normalized scattering power, defined as $P_{sc}\frac{k_0}{2}\sqrt{\frac{\mu_0}{\epsilon_0}}$.
Here, we set a derivation in the amplitude $E_i$ and incident angle $\Phi_i$ from the illumination beams, for the $i$th control wave, $i = 3, 4, 5, 6$ in (a-d), respectively. Each contour plot reveals a wide region to support invisible cloak with toleration. }
\end{figure}

\section{Appendix C: Robustness of interferometric coherent waves}
\begin{figure*}
\includegraphics[width=14.5cm]{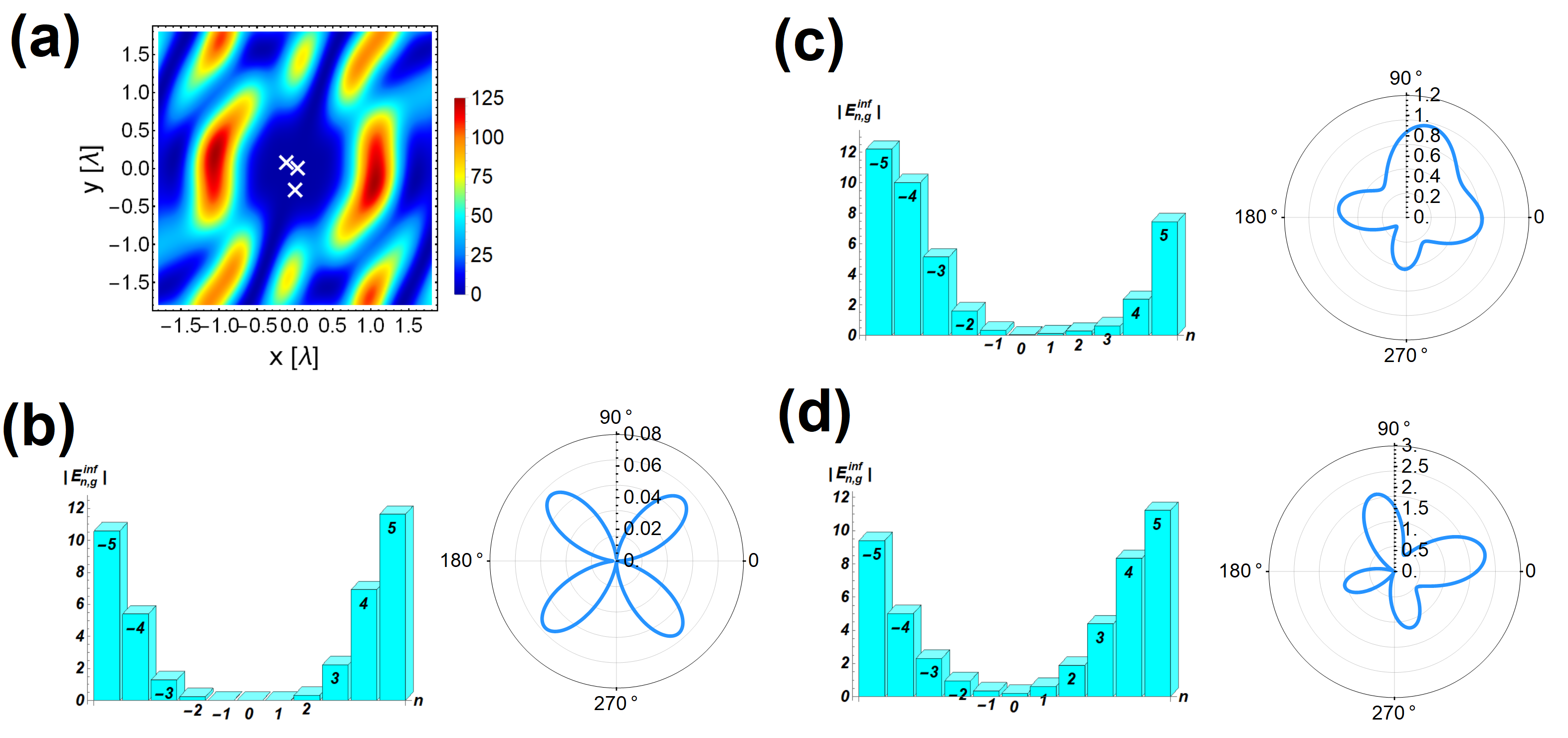}
\caption{(a) Three locations marked by White-color are used to study the offset in displacement on the scatterer. Here, the background fields is generated with the same illumination setting give in  Fig. 2 (d) of our manuscript. 
The corresponding general interfering factors and far-field scattering pattern are depicted in (b-d).}
\end{figure*}

In Fig. 6, we analyze the normalized scattering powers, defined as $P_{sc}\sqrt{\mu_0/\epsilon_0}k_0/2$, with a deviation in the amplitude and incident angles of illumination waves.

We find that for the control waves $E_3$, $E_4$, and $E_5$, as shown in Figs.  6(a)-6(c), the supporting invisible cloak is almost insensitive to the variations in amplitudes.
Instead, the deviation in phase gives a larger change.
For the control wave $E_6$, the opposite situation is observed from Fig. 6(d).
Nevertheless, in terms of the variation on illumination wave, both for amplitude and angles, the advantageous of our interferometric coherent wave certainly can endure the mismatching.

In addition to the derivation in the illumination waves, we also investigate the offset in displacement.
Here, three different locations are arbitrarily studied  for the scatterer, see the marked points in Fig. 7(a).
The corresponding generalized interfering factors and far-field scattering pattern are depicted in (b) for the location at $(b=0.075\lambda,\beta=\pi/4)$, (c)  $(b=0.225\lambda,\beta=3\pi/4)$, and (d)  $(b=0.35\lambda,\beta=3\pi/2)$, respectively.
One can see clearly that even with  dislocations, the target scattering events at  $n = 0, \pm 1, \pm 2$ are all still suppressed, which demonstrate the robustness of invisibility.


\begin{thebibliography}{99}


\bibitem{super1}
S. Arslanagi\'c and R. W. Ziolkowski,
``Highly Subwavelength, Superdirective Cylindrical Nanoantenna,''
Phys. Rev. Lett. \textbf{120}, 237401 (2018).


\bibitem{super2}
W. Liu, A. E. Miroshnickenko, D. N. Neshev, and Y. S. Kivshar, 
``Broadband unidirectional scattering by magnetoelectric
core–shell nanoparticles,'' ACS Nano \textbf{6}, 5489 (2012).

\bibitem{super3}
I. M. Hancu, A. G. Curto, M. Castro-Lopez, M. Kuttge, and
N. F. van Hulst, 
``Multipolar interference for directed light emission,'' Nano Lett. \textbf{14}, 166 (2014).


\bibitem{absorber1}
H. Noh, Y.-D. Chong, A. D. Stone, and H. Cao,
``Perfect coupling of light to surface plasmons by coherent absorption,''
Phys. Rev. Lett. \textbf{108}, 186805 (2012).


\bibitem{absorber2}
P. Bai, Y. Wu, and Y. Lai, 
``Multi-channel coherent perfect absorbers,'' 
Europhys. Lett. \textbf{114}, 28003 (2016).




\bibitem{magnetic1}
J. A. Schuller, R. Zia, T. Taubner, and M. L. Brongersma,
``Dielectric Metamaterials Based on Electric and Magnetic Resonances of Silicon Carbide Particles,''
Phys. Rev. Lett. \textbf{99}, 107401 (2007).

\bibitem{magnetic2}
T. Feng, Y. Xu, W. Zhang, and A. E. Miroshnichenko,
``Ideal Magnetic Dipole Scattering,''
Phys. Rev. Lett. \textbf{118}, 173901 (2017).

\bibitem{magnetic3}
W. Liu,
``Generalized Magnetic Mirrors,''
Phys. Rev. Lett. \textbf{119}, 123902 (2017).





\bibitem{kerker1}
J. Y. Lee, A. E. Miroshnichenko, and R.-K. Lee, 
``Reexamination of Kerker’s conditions by means of the phase diagram,'' 
Phys. Rev. A \textbf{96}, 043846 (2017).

\bibitem{kerker2}
J. Y. Lee, A. E. Miroshnichenko, and R.-K. Lee, 
``Simultaneously nearly zero forward and nearly zero backward scattering objects,''
Opt. Express \textbf{26},  30393 (2018).

\bibitem{kerker3}
W. Liu and Y. S. Kivshar, ``Generalized Kerker effects in nanophotonics
and meta-optics,'' Opt. Express \textbf{26}, 13085 (2018).






\bibitem{anapole}
A. E. Miroshnichenko, A. B. Evlyukhin, Y. F. Yu, R. M. Bakker, A. Chipouline,
A. I. Kuznetsov, B. Luk\'yanchuk, B. N. Chichkov, and Y. S. Kivshar,
``Nonradiating anapole modes in dielectric nanoparticles,''
Nat. Commun. \textbf{6}, 8069 (2015).


\bibitem{superscattering1}
Z. Ruan and S.-H. Fan,
``Superscattering of Light from Subwavelength Nanostructures,''
Phys. Rev. Lett. \textbf{105}, 013901 (2010).

\bibitem{superscattering2}
C. Qian, X. Lin, Y. Yang,  X. Xiong, H. Wang, E. Li,
I. Kaminer, B. Zhang, and H. Chen,
``Experimental Observation of Superscattering,''
Phys. Rev. Lett. \textbf{122}, 063901 (2019).


\bibitem{controlling}
J. B. Pendry, D. Schurig, and D. R. Smith, ``Controlling electromagnetic fields," Science \textbf{312}, 1780 (2006).

\bibitem{conformal} 
U. Leonhardt,
``Optical conformal mapping,"
 Science \textbf{312}, 1777 (2006).
 

 
 \bibitem{alu1}
 A. Al\'u and N. Engheta, 
 ``Achieving transparency with plasmonic and metamaterial coatings,''
 Phys. Rev. E \textbf{72}, 016623
(2005).
 
 \bibitem{alu2}
 A. Al\'u and N. Engheta, 
 ``Plasmonic materials in transparency and cloaking problems: mechanism, robustness, and physical insights,''
 Opt. Express \textbf{15}, 3318 (2007).

\bibitem{kerker} 
M. Kerker, 
``Invisible bodies,''
J. Opt. Soc. Am. \textbf{65}, 376 (1975).


\bibitem{acoustic1}
S. A. Cummer and D. Schurig,
``One path to acoustic cloaking,'
New J. Phys. \textbf{9}, 45 (2007).

\bibitem{acoustic2}
S. Zhang, C. Xia, and N. Fang,
``Broadband acoustic cloak for ultrasound waves,''
Phys. Rev. Lett. \textbf{106}, 024301 (2011).

\bibitem{acoustic3}
S. A. Cummer, B.-I. Popa, D. Schurig, D. R. Smith, J. Pendry,
M. Rahm, and A. Starr,
``Scattering theory derivation of a 3D acoustic cloaking shell,''
Phys. Rev. Lett. \textbf{100}, 024301 (2008).



 \bibitem{sc1}
 M. D. Guild, A. Al\'u, and M. R. Habermann, 
 ``Cancellation of acoustic scattering from an elastic sphere,'' J. Acoust.
Soc. Am. \textbf{129}, 1355 (2011).
 
 \bibitem{sc2}
 L. Sanchis,  V. M. Garcı\'a-Chocano,  R. Llopis-Pontiveros,  A. Climente, 
J. Martı\'nez-Pastor,  F. Cervera,  and J. Sa\'nchez-Dehesa,
 ``Three-dimensional axisymmetric cloak based on the cancellation
of acoustic scattering from a sphere,''
Phys. Rev. Lett. \textbf{110}, 124301 (2013).





\bibitem{fluid1}
M. Farhat, S. Enoch, S. Guenneau, and A. B. Movchan,
``Broadband cylindrical acoustic cloak for linear surface waves in a fluid,'' Phys. Rev. Lett. \textbf{101}, 134501 (2008).

\bibitem{fluid2}
Y. A. Urzhumov and D. R. Smith,
``Fluid flow control with transformation media,''
Phys. Rev. Lett. \textbf{107}, 074501 (2011).

\bibitem{thermal1}
R. Schittny, M. Kadic, S. Guenneau, and M. Wegener,
``Experiments on transformation thermodynamics: molding the flow of heat,'' Phys. Rev. Lett. \textbf{110}, 195901 (2013).

\bibitem{thermal2}
S. Guenneau, C. Amra, and D. Veynante,
``Transformation thermodynamics:
cloaking and concentrating heat flux,'' Opt. Express \textbf{20}, 8207 (2012).

\bibitem{sc8} 
M. Farhat, P.-Y. Chen, H. Bagci, C. Amra, S. Guenneau, and A. Al\'u,
``Thermal invisibility based on scattering
cancellation and mantle cloaking,''
Sci. Rep. \textbf{5}, 9876 (2015).

\bibitem{quantum1}
S. Zhang, D. A. Genov, C. Sun, and X. Zhang,
``Cloaking of matter waves,''
Phys. Rev. Lett. \textbf{100}, 123002 (2008).

\bibitem{quantum2}
D.-H. Lin,
``Cloaking spin-1/2 matter waves,''
Phys. Rev. A \textbf{81}, 063640 (2010).

\bibitem{quantum3}
 D.-H. Lin,
 ``Cloaking two-dimensional fermions,''
 Phys. Rev. A \textbf{84}, 033624 (2011).
 
 
 


\bibitem{sc6}
J. Y. Lee  and R.-K. Lee,
``Hiding the interior region of core-shell nanoparticles with quantum invisible cloaks,''
Phys. Rev. B \textbf{89}, 155425 (2014).

\bibitem{sc7}
B. Liao, M. Zebarjadi, K. Esfarjani, and G. Chen,
``Cloaking core-shell nanoparticles from conducting electrons in solids,''
Phys. Rev. Lett. \textbf{109}, 126806 (2012).


 
 \bibitem{elastic1}
 N. Stenger, M. Wilhelm, and M. Wegener,
 ``Experiments on elastic cloaking in thin plates,'' Phys. Rev. Lett. 
\textbf{108}, 014301 (2012).
 
 \bibitem{elastic2}
 M. Farhat, S. Guenneau, and S. Enoch,
 ``Ultrabroadband elastic cloaking in thin plates,'' Phys. Rev. Lett. 
\textbf{103}, 024301 (2009).

\bibitem{elastic3}
M. Farhat, S. Guenneau, S. Enoch, and A. B. Movchan,
``Cloaking bending waves propagating in thin elastic plates,''
Phys. Rev. Lett. \textbf{79},  033102 (2009).
 

 

\bibitem{carpet}
J. Li and J. B. Pendry,
``Hiding under the carpet: a new strategy for cloaking,''
Phys. Rev. Lett \textbf{101}, 203901 (2008).



\bibitem{quest1}
D. A. B. Miller,
``On perfect cloaking,''
Opt. Express \textbf{14}, 12457 (2006).


\bibitem{quest2}
B. Zhang,
``Electrodynamics of transformation-based invisibility cloaking,''
Light Sci. Appl. \textbf{1}, e32 (2012).

\bibitem{quest3}
R. Fleury, F. Monticone, and A Al\'u,
``Invisibility and cloaking: origins, present, and future perspectives,''
Phys. Rev. Appl. \textbf{4}, 037001 (2015).

\bibitem{quest4}
F. Monticone and A. Al\'u,
``Do cloaked objects really scatter less ?'' Phys. Rev. X \textbf{3}, 041005 (2013)


\bibitem{sum}
E. M. Purcell, 
``On the Absorption and Emission of Light by
Interstellar Grains,'' Astrophys. J. \textbf{158}, 433 (1969).


\bibitem{linear1}
J. Y. Lee, Y.-H. Chung, A. E. Miroshnichenko, and R.-K. Lee,
``Linear control of light scattering with
multiple coherent waves excitation,'' 
Opt. Lett. {\bf 44}, 5310 (2019).



\bibitem{book1}
C. F. Bohren and D. R. Huffman, \textit{Absorption and Scattering of Light by Small Particles} (Wiley 1983).

\bibitem{handbook}
E. D. Palik, \textit{Handbook of Optical Constants of Solids}
(Academic Press, 1985).


\bibitem{linear2}
Z. Xi and H. P. Urbach,
``Magnetic Dipole Scattering from Metallic Nanowire for Ultrasensitive Deflection Sensing,'' Phys. Rev. Lett. \textbf{119}, 053902 (2017).

\bibitem{linear3}
T. Das, P. P. Iyer, R. A. DeCrescent, and J. A. Schuller,
``Beam engineering for selective and enhanced coupling to multipolar resonances,''
Phys. Rev. B \textbf{92}, 241110(R) (2015).


\bibitem{linear4}
L. Wei, A. V. Zayats, and F. J. Rodriguez-Fortu\~no ,
``Interferometric Evanescent Wave Excitation of a Nanoantenna for Ultrasensitive Displacement and Phase Metrology,''
Phys. Rev. Lett. \textbf{121}, 193901 (2018).














\bibitem{addition}
P. T. Kristensen, P. Lodahl, and J. Mørk,
``Light propagation in finite-sized photonic crystals: multiple scattering using an electric field integral equation,''
J. Opt. Soc. Am. B \textbf{27}, 228 (2010).

\bibitem{comsol}
COMSOL Multiphysics, http://www.comsol.com/.


\end{thebibliography}
\end{document}